\documentclass[twocolumn,pdflatex,sn-nature]{sn-jnl}
\usepackage{graphicx}%
\usepackage{hyperref}
\usepackage{xcolor}%
\usepackage{multirow}%
\usepackage{amsmath,amssymb,amsfonts}%
\usepackage{amsthm}%
\usepackage{mathrsfs}%
\usepackage[title]{appendix}%
\usepackage{textcomp}%
\usepackage{manyfoot}%
\usepackage{booktabs}%
\usepackage{algorithm}%
\usepackage{algorithmicx}%
\usepackage{algpseudocode}%
\usepackage{listings}%
\theoremstyle{thmstyleone}%
\theoremstyle{thmstyletwo}%
\theoremstyle{thmstylethree}%
\begin{document}
\title[Roton-Based Unified Origin of Low-Temperature Anomalies in SrTiO$_3$]{Roton-Based Unified Origin of Low-Temperature Anomalies in SrTiO$_3$}

\author*[1]{\fnm{Fernando} \sur{G\'omez-Ortiz}}\email{fgomez@uliege.be}

\author[1]{\fnm{Louis} \sur{Bastogne}}

\author*[1]{\fnm{Philippe} \sur{Ghosez}}\email{Philippe.Ghosez@uliege.be}

\affil*[1]{\orgdiv{Theoretical Materials Physics, Q-MAT}, \orgname{Universit\'e de Li\`ege}, \orgaddress{\street{} \postcode{B-4000} \city{Li\`ege}, \country{Belgium}}}

\abstract{Strontium titanate, SrTiO$_3$, is known as an intriguing and fascinating compound hosting a remarkable range of low-temperature quantum phenomena, from quantum paraelectricity~\cite{Cowley-62,Muller-79}, to anomalous lattice dynamics~\cite{Muller-91,Scott-97} and, when doped, superconductivity at exceptionally low carrier densities~\cite{Schooley-64}. Despite decades of intense debate, the potential link between these unusual features remains elusive. 
Here we show that, at low temperatures, bulk SrTiO$_3$ supports an intrinsic acousto-roto-polar composite structural instability, driven by anharmonic coupling between acoustic, oxygen-octahedra-rotation and polar phonons. A roton-like minimum develops in the transverse acoustic branch at finite wavevector, and gives rise to a coherent modulated state with a characteristic wavelength of 14-18 unit cells that reconciles seemingly disparate experimental observations accumulated over decades. 
Its composite character provides together a natural explanation for the long-standing Müller anomaly~\cite{Muller-91} and for the appearance of polar nanoregions~\cite{Zhang-26}. Moreover, its long wavelength modulated character is responsible for the opening of a gap close to the conduction band minimum, providing a new microscopic channel for electron–phonon pairing and superconductivity at ultra-low carrier densities. These results establish the emergence of a roton-type acousto-roto-polar structural modulation as the unifying origin of the low-temperature physics of SrTiO$_3$, bridging its anomalous lattice dynamics to its electronic and superconducting properties.} 


\maketitle

For more than six decades, SrTiO$_3$ has continued to surprise and capture attention, gradually becoming an iconic material in condensed matter physics.
Still today, it remains a remarkable platform for materials science, serving as a foundational substrate for epitaxial growth and strain engineering~\cite{Schlom-08}, while also offering rich interfacial physics such as the emergence of a superconducting two-dimensional electron gas at the  SrTiO$_3$/LaAlO$_3$ interface~\cite{Ohtomo2004,Reyren-07}, to name just a few examples.
Much of the historical interest for SrTiO$_3$ stems from its remarkable proximity to a ferroelectric instability~\cite{Cowley-62,Muller-79} together with a range of low-temperature phenomena whose microscopic origins remain debated~\cite{Muller-91,Schooley-64,Collignon-19}. Combined with its structural simplicity, this rich physics positions SrTiO$_3$ as a prototypical system for exploring the interplay between lattice fluctuations, structural distortions and electronic properties~\cite{Kleemann-20}.

The first indication that SrTiO$_3$ defies the conventional picture of a displacive ferroelectric~\cite{Cowley-62,Muller-79} came from its lattice dynamics. 
At high temperatures the material crystallizes in the cubic perovskite structure, undergoing at 105 K the well-known antiferrodistortive transition~\cite{Muller-58}, characterized by alternating oxygen octahedra rotations (OOR)~\cite{Scott}. 
Below this transition, a transverse optical ferroelectric mode continues to soften upon cooling~\cite{Cowley-62,Yamada-69,Viana-94}. In an ordinary displacive ferroelectric, this behavior would culminate in a polar instability. Instead, the softening saturates at low temperature and long-range ferroelectric order never develops, giving rise to the celebrated quantum paraelectric ground state~\cite{Muller-79,Viana-94}. This unusual behavior is  indeed commonly attributed to quantum fluctuations~\cite{Zhong-96,Rowley2014}, leaving SrTiO$_3$ at the verge of a ferroelectric instability and exceptionally sensitive to perturbations such as strain~\cite{Haeni-04}, isotopic substitution~\cite{Itoh-99} or lattice expansion~\cite{Uwe-76,Bethe-71,Cavalleri-19}.

Rather than resolving the puzzle, increasingly detailed studies of the low-temperature behavior of SrTiO$_3$ have given rise to a succession of new mysteries.
High-resolution electron paramagnetic resonance experiments revealed the so-called Müller anomaly~\cite{Muller-91}: rather than evolving monotonically toward saturation below the 105 K transition, the antiferrodistortive OOR angle exhibits an unexpected enhancement at low temperature. The microscopic origin of this enhancement  is unclear  and was tentatively attributed to several mechanisms, including collective quantum states~\cite{Muller-91}, dynamical anharmonic coupling~\cite{Scott-99,Scott-97}, polar clusters, the onset of second sound, domain-related effects, as well as a possible role of defects~\cite{Courtens-96,Annette-24}. Whether the anomaly reflects an intrinsic collective instability of the lattice or instead originates from local, domain-related, dynamical or extrinsic effects remains an open question.

Another challenge concerns superconductivity. SrTiO$_3$ becomes superconducting under doping at extraordinarily low carrier densities~\cite{Schooley-64,Koonce-67,Lin-13,Lin-14,Collignon-19}, far below those characteristic of conventional superconductors. Despite extensive experimental and theoretical efforts, the microscopic origin of superconductivity remains unsettled, with competing proposals concerning the character of the superconducting order, the interactions responsible for pairing, or more fundamentally, the spatial distribution of carriers within the material~\cite{Gastiasoro-20,Triscone-20,Szot-24}.

These longstanding puzzles have recently gained renewed interest following direct imaging of previously anticipated finite-wavevector polar order~\cite{Uwe-89,Salje-13}.
Polar nano-regions have been observed in SrTiO$_3$~\cite{Fauque-22,Zhang-26}, while long-wavelength polar modulations have been stabilized by optical excitation~\cite{Orenstein-25,Wang-26} or uniaxial strain~\cite{Yajun-26}. Yet, the extent to which these finite-wavevector polar states represent an intrinsic instability of SrTiO$_3$ and are connected to its anomalous low-temperature lattice and electronic properties, remains unknown.

In the literature, the Müller anomaly, finite-wavevector polar order, and superconductivity have generally been regarded as distinct low-temperature phenomena emerging in the same material. As a consequence, theoretical descriptions have largely treated these phenomena within separate frameworks. Whether these disparate observations originate from a single underlying principle has remained elusive. But, could eventually the apparently diverse low-temperature anomalies of SrTiO$_3$ be distinct manifestations of a common finite-wavevector lattice instability?

To address this question, we combine state-of-the-art first-principles calculations with accurate second-principles simulations~\cite{Ghosez-22}, relying on exhaustively-validated effective interatomic potentials that enable predictive fully atomistic investigation of anharmonic lattice dynamics at finite temperature in large supercells, without empirical parameters (see Method section). We consider unusual supercells allowing to track the appearance of long-wavelength atomic modulations and to explore their various consequences on the structural, electronic and dynamical properties. Our investigations highlight intrinsic anharmonic lattice couplings that give rise to a modulated state, which then provides a unified framework for understanding the anomalous low-temperature behavior of SrTiO$_3$ and connects phenomena that have remained seemingly unrelated for more than half a century.
\begin{figure*}[h]
     \centering
      \includegraphics[width=\textwidth]{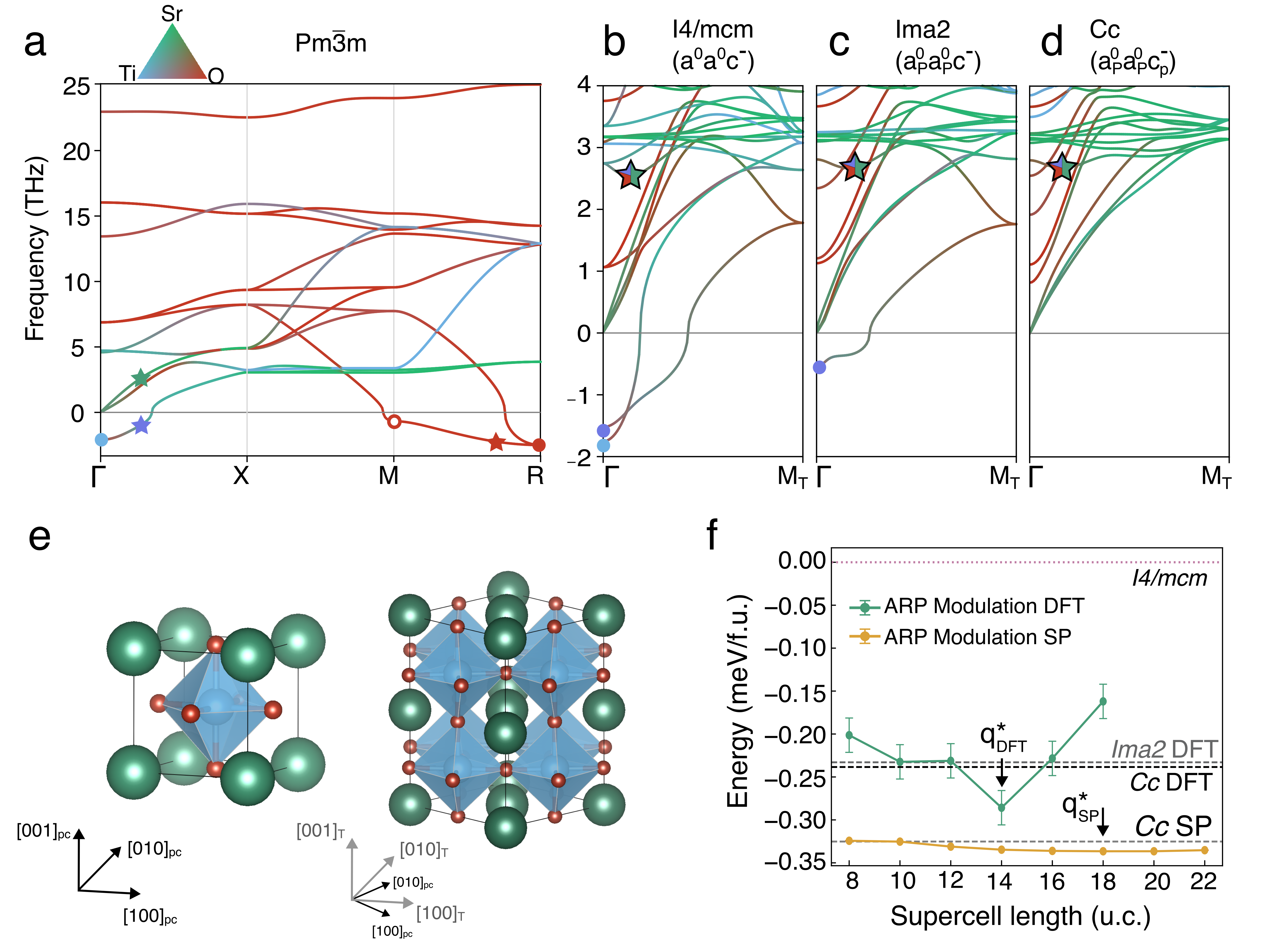}
      \caption{\textbf{Phonon instabilities and energetics of SrTiO$_3$.} Phonon dispersion curves of SrTiO$_3$ at $T=0$ K in the \textbf{a,} $Pm\bar{3}m$, \textbf{b,} $I4/mcm$, \textbf{c,} $Ima2$ and \textbf{d,} $Cc$ phases, obtained from density functional perturbation theory. Notice that, as depicted in \textbf{e,} phonons in the last three panels are computed in the 20-atom tetragonal cell, which is rotated by 45$^\circ$ with respect to the cubic reference; in this setting, $M_T=(0.5,0.5,0)$ corresponds to pseudocubic $X$-point in panel {\bf a}. Blue and red points in \textbf{a} highlight respectively polar and oxygen octahedra rotation instabilities whereas stars represent the modes present in the acousto-roto-polar (ARP) modulated state. Blue points in \textbf{b} represent the splitting between the in-plane (light blue) and out-of-plane (dark blue) polar modes. Mixed-color stars in \textbf{b}, \textbf{c} and \textbf{d} indicate the mixed character of the ARP phonon branch. \textbf{f,} Energy lowering (after full relaxation) resulting from the  condensation of the ARP modulation in the $I4/mcm$ phase, as a function of its wavelength $\lambda$, obtained using DFT (green) and second-principles  (yellow) calculations. Black arrows locate the position of the energy minima. DFT error bars reflect the use of slightly different effective $k$-point meshes, resulting from the different supercell sizes.} 
      \label{fig:instability_phonons} 
\end{figure*}

We first revisit step-by-step the phase transition sequence of SrTiO$_3$ under cooling. At room temperature, it crystallizes in the prototypical cubic perovskite structure. However, as shown in Fig.~\ref{fig:instability_phonons}(a), this cubic $Pm\bar{3}m$ phase exhibits antiferrodistortive instabilities at $R$-point (filled red dot) and $M$-point (open red dot), associated respectively to anti-phase and in-phase OOR, as well as polar instabilities at the zone center (blue dot). On cooling, SrTiO$_3$ undergoes at 105 K an antiferrodistortive transition to a tetragonal $I4/mcm$ structure~\cite{Muller-58}, arising from the condensation of $R$-point OOR along $z$ ($a^0a^0c^-$ in Glazer's notations~\cite{Glazer-72}),  with amplitude $\phi_z^R$~\cite{Yamada-69,Scott}. 

As shown in Fig.~\ref{fig:instability_phonons}b, polar modes remain unstable in the tetragonal phase but the presence of  $\phi_z^R$ lifts the degeneracy, favoring the polar distortions in the $xy$-plane (light blue dot) respect to that along $z$ (dark blue dot). The condensation of in-plane polar modes in the $I4/mcm$ structure lowers the energy and produces an $Ima2$ structure with polarization along the $[110]_{\rm pc}$ direction ($P_x^0=P_y^0$). In the latter, the polar mode along $z$ remains unstable (Fig.~\ref{fig:instability_phonons}c) and its additional condensation further lowers marginally 
the energy and produces a $Cc$ phase with $p_z^0 < P_x^0=P_y^0$. Experimentally, these polar condensations never happen: the polar instabilities are renormalized by quantum fluctuations and SrTiO$_3$ appears as a quantum paraelectric~\cite{Muller-79,Zhong-96} with $I4/mcm$ structure. However, the $Cc$ phase has a lower internal energy and, as shown in Fig.~\ref{fig:instability_phonons}d, does not show any unstable phonon mode so that it can be seen as the classical thermodynamical ground state of SrTiO$_3$.

Interestingly, we can see in Fig.~\ref{fig:instability_phonons}b-d  a previously overlooked roton-like feature (star), consisting in a local minimum at a finite wavevector $q^\star$, away from high-symmetry points, in the folded acoustic phonon branch of the $I4/mcm$ and related $Ima2$ and $Cc$ phases. 
While this feature is presented here along the $\Gamma-M_T$ direction, similar features are also present along other in-plane directions, including $\Gamma-X_T$ (see Supplementary Information) but not along the tetragonal direction $\Gamma-Z_T$.
This mode, further labeled as the {\it roton}, is dynamically stable. However, as highligthed in Fig.~\ref{fig:instability_phonons}f, its finite condensation in the $Ima2$ phase --as well as that of nearby phonons at close $q$ in the same branch-- produces, after full atomic relaxation, an unexpected stable modulated state located at an energy eventually lower than that of the $Cc$ phase. 
This establishes that the real thermodynamical ground state of SrTiO$_3$ is not the usually expected $Cc$ phase but rather a modulated structure. 

Strikingly, this unexpected  ground state adopts the modulation associated with the roton wavevector $q^\star$.
To confirm this correspondence, we condense the roton-like modes at slightly different wavevectors, generating modulated structures with distinct wavelengths and corresponding supercell lengths (Fig.~\ref{fig:instability_phonons}f). After full atomic relaxation, these structures remain stable over a range of modulation wavelengths while preserving the characteristic distortions of the roton eigenvector. Their relative energies vary with the chosen wavevector, with the lowest-energy structure corresponding to the roton minimum at $q^\star$ (Fig.~\ref{fig:instability_phonons}f). This demonstrates that both the character of the distortion and the equilibrium modulation length are selected by the roton lattice feature.

\begin{figure*}[h]
     \centering
      \includegraphics[width=\textwidth]{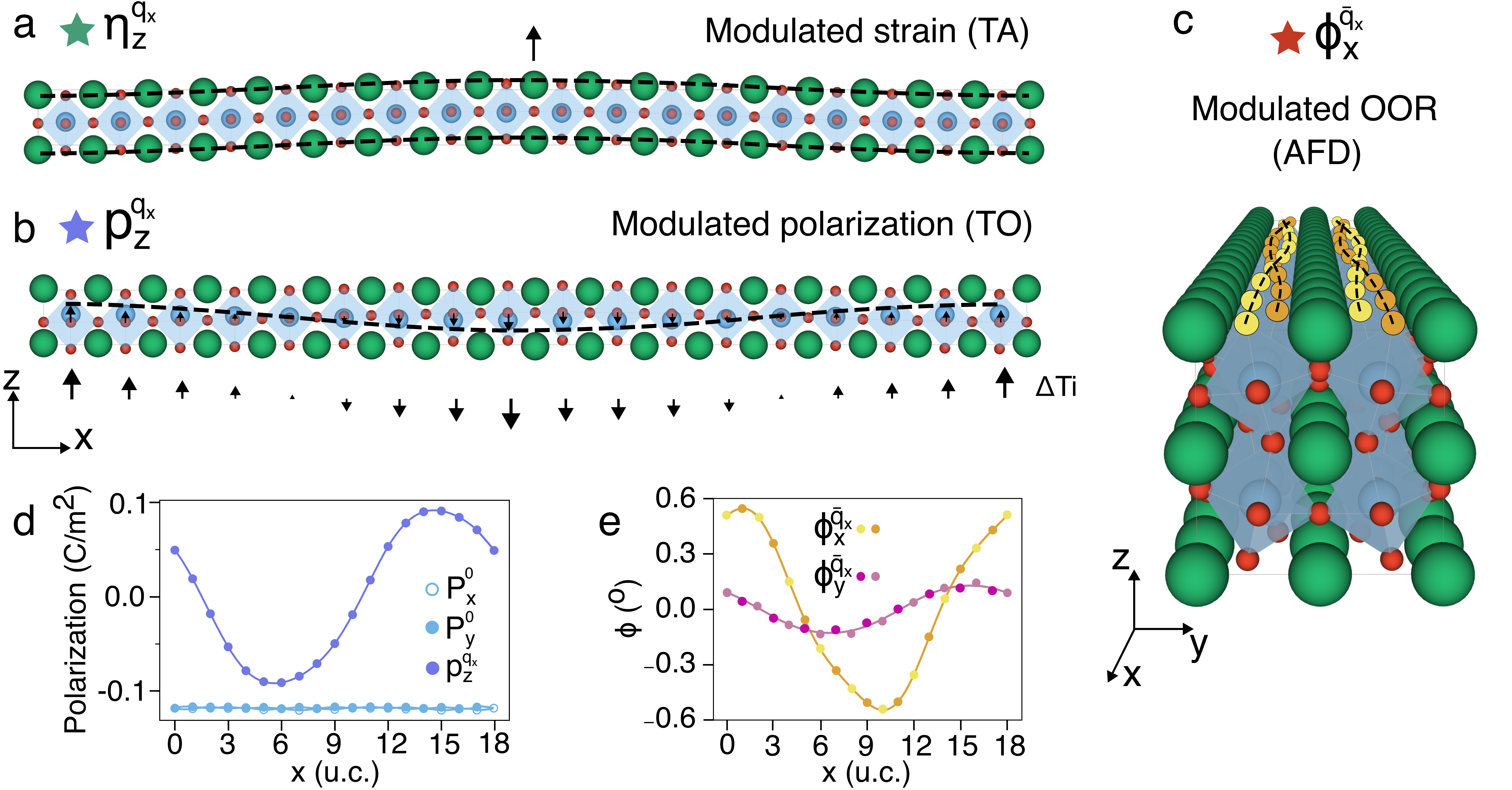}
      \caption{\textbf{Characteristic modes composing the modulated atomic distortion.} 
      \textbf{a,} Transverse-acoustic mode $\eta_z^{q_x}$; \textbf{b,} Transverse-optic mode $p_z^{q_x}$, with atomic displacements along $z$; \textbf{c,} Modulated OOR mode about the $x$ axis $\phi_x^{\bar{q}_x}$. Sr, O, and Ti atoms are shown in green, red, and blue, respectively. For clarity, the amplitudes of the distortions are exaggerated.
      \textbf{d,} Modulated ($p_z^{q_x}$, dark blue) and constant in-plane ($P_x^0= P_y^0$, light blue) polarization components characterizing the acousto-roto-polar modulated ground state at $T=0$ K, as obtained after full DFT relaxation. \textbf{e,} Modulated OOR about the $x$-axis (yellow/orange) and $y$-axis (magenta/purple), characterizing the acousto-roto-polar modulated ground state at $T=0$ K, as obtained after full DFT relaxation. Since the OOR alternate in sign in consecutive unit cells $i$ along the $x$ direction, a phase factor $(-1)^i$ has been applied to the plotted profiles and slightly different colors are used to distinguish the two AFD sublattices.} 
      \label{fig:ModesPDW} 
\end{figure*}

To get better insight into the roton mode yielding the modulated ground state, we decompose its eigenvector in terms of the modes of the high-symmetry cubic $Pm\bar{3}m$ phase. We see that it arises from the combination of 3 main modes (green, blue and red stars in Fig.~\ref{fig:instability_phonons}a): (i) a transverse acoustic mode $\eta_z^{q_x}$ at finite wavevector $q_x=(\tfrac{1}{\lambda},0,0)$, associated to a long-wavelength strain modulation along $x$, with atomic displacements along $z$ (green star, Fig.~\ref{fig:ModesPDW}a); (ii) a transverse optical mode  $p_z^{q_x}$ at the same $q_x$, associated to a long-wavelength modulation along $x$ of the polarization component along $z$ (blue star, Fig.~\ref{fig:ModesPDW}b); and (iii) a two-component AFD mode ($\phi_x^{\bar{{q}_x}},~\phi_y^{\bar{{q}_x}}$) at $\bar{q}_x=(\tfrac{1}{2}-\tfrac{1}{\lambda},\tfrac{1}{2},\tfrac{1}{2})$, combining long-wavelength distinct modulations along $x$ of OOR about the $x$- and $y$-axes respectively (red star, Fig.~\ref{fig:ModesPDW}c).
As such, the roton at $q^\star$ appears therefore as a composite acousto-roto-polar (ARP) mode combining acoustic (50\% $\eta_z^{q_x}$), OOR (25\% $\phi_x^{\bar{{q}_x}},~\phi_y^{\bar{{q}_x}}$) and polar (25\% $p_z^{q_x}$)  modulations (see Supplementary Video 1). The resulting ground state is closely related to the $Cc$ phase (condensing $\phi_z^R$, $p_z^0<P_x^0=P_y^0)$ but with the homogeneous polarization $p_z^0$ along the principal rotation axis replaced by the roton ARP distortion in which the polarization along $z$ is modulated along a perpendicular direction, here $x$ ($p_z^{q_x}$, Fig.~\ref{fig:ModesPDW}d).  At the same time, the octahedral rotations about the two transverse axes acquire a corresponding modulation with the same wavelength (Fig.~\ref{fig:ModesPDW}e).  

Remarkably, although long-wavelength polar modulations have recently been stabilized experimentally in SrTiO$_3$ by optical excitation or uniaxial tensile strain~\cite{Orenstein-25,Wang-26}, the modulated state presented here emerges spontaneously upon unconstrained structural relaxation. 
Moreover, as shown in Fig.~\ref{fig:nanoregions}, condensation of ARP roton modes along different directions leads, after full structural relaxation, to spatially modulated polar nanoregions whose morphology and length scale are consistent with those recently imaged experimentally~\cite{Zhang-26}. Together, this highlights that the  recently observed long-wavelength modulated polar order is likely to have a composite ARP character and is not simply an externally induced state, but also an intrinsic instability of SrTiO$_3$.

The emergence of such a low-energy  modulated state must be found in the energetics of the relevant structural distortions: since the roton mode is dynamically stable, the modulated state must be stabilized by cooperative anharmonic interactions between the underlying $p_z^{q_x},~\eta_z^{q_x}$ and $\phi^{\bar{{q_x}}}$ modes. Similar anharmonic couplings between modulated polar, acoustic, and antiferrodistortive modes have been recently highlighted to drive instabilities in strained SrTiO$_3$~\cite{Yajun-26}. Here, both second-principles and full first-principles relaxations converge to the same modulated structure, showing that these couplings are sufficiently strong even in unstrained SrTiO$_3$ to stabilize a hidden phase, despite the absence of a corresponding harmonic instability (Fig.~\ref{fig:instability_phonons}f).

Roto-polar~\cite{Sun-26} and acousto-polar~\cite{Guzman-23,Muller-91,Orenstein-25} couplings have both been invoked separately in the literature as possible explanations for a finite-$q^\star$ minimum in the phonon branches of SrTiO$_3$. 
Our fully atomistic simulations reconcile these two pictures by revealing that the relevant acoustic branch has both substantial polar and rotational characters. Thus, a joint ARP coupling contribute to the collective instability generating the finite-$q^\star$ softening of the hybridized branch in the tetragonal phase and with neither interaction alone (roto-polar or acousto-polar) fully capturing the character of the energy-lowering mode. 

Such an ARP coupling was in fact anticipated by Scott and Ledbetter~\cite{Scott-97}, who pointed out that it is allowed by symmetry (see Supplementary Information). However, symmetry allowance does not necessarily imply that the coupling is realized, as the modes must also be sufficiently close in energy to interact appreciably.
Indeed, temperature-induced changes in the relative energies of the modes can alter their energetic proximity and, as we show below, this temperature dependence can substantially modify the ARP roton, which is not necessarily preserved at all temperatures.
\begin{figure*}[h]
     \centering
      \includegraphics[width=\textwidth]{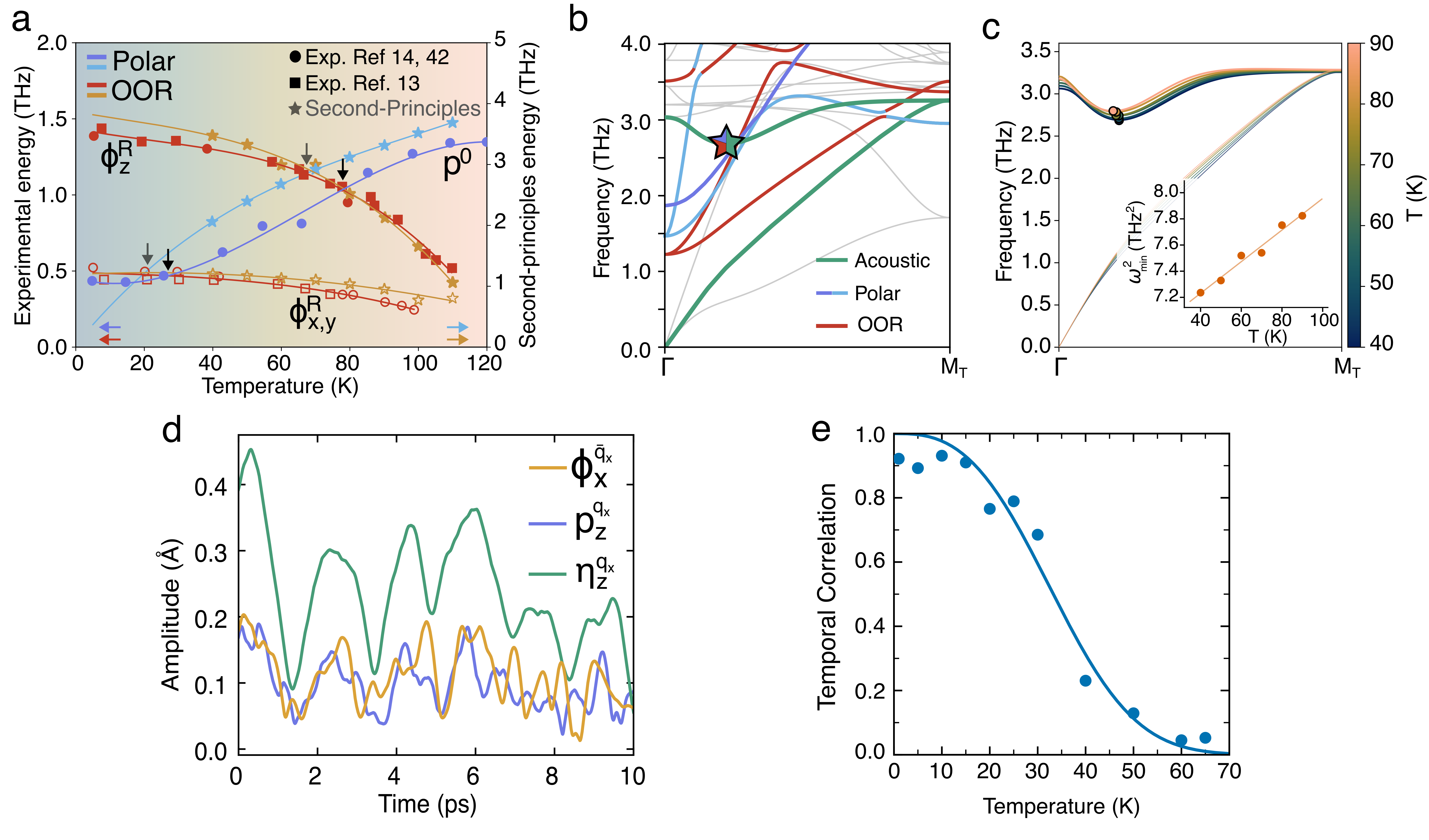}
      \caption{\textbf{Finite temperature behavior.} \textbf{a,} Temperature evolution of the frequency of (i) the transverse-optic polar mode $P_{x,y}^0$ (experiment: dark blue; second principles: light blue) and (ii) the in-plane ($\phi_{x,y}^R$, open symbols) and tetragonal axis ($\phi_z^R$, filled symbols)  OOR modes (experiment: red; second principles: orange). Dots, squares denote experimental data from Refs.~\cite{Yamada-69,Yamada-2,Scott}, while stars represent the temperature-renormalized second-principles frequencies. Black and grey arrows indicate the temperatures where crossing between polar and OOR frequencies occur. \textbf{b,} Temperature-renormalized phonon dispersion curves of SrTiO$_3$ at $T=30$ K in the $I4/mcm$ phase, obtained from second-principles anharmonic dynamics. The acoustic branch presenting the roton, the polar branches and the OOR branches are colored as indicated in the legend. \textbf{c,} Temperature evolution of the folded acoustic branch in the in the $I4/mcm$ phase presenting the roton minimum, with the softening of the latter as an inset. 
      \textbf{d,} Time evolution of the different distortion modes during the dynamics at $T=20$ K, as predicted by second-principles simulations. \textbf{e,} Temporal correlation between the $\phi_x^{\bar{{q}_x}}$, $\eta_z^{q_x}$ and $p_z^{q_x}$ modes, obtained from the dynamics, as the system evolves from a coherent to an incoherent regime as a function of temperature. The blue curve is a guide to the eye showing a exponential-like decay.} 
      \label{fig:Temperature} 
\end{figure*}

Our second-principles temperature-dependent phonon results reveal two  distinct temperature regimes in Fig.~\ref{fig:Temperature}(a).
Above $T \approx 80$ K, the polar modes are at much higher frequencies than the in-plane antiferrodistortive modes, becoming instead comparable in frequency to the harder out-of-plane rotational modes. Upon cooling, the polar mode softens, while the antiferrodistortive branches harden, bringing both, in-plane AFD and polar branches, closer to the acoustic branches at $T\lesssim 30$K. As a result, the energetic proximity between the modes in this regime favors the coupling which was precluded at $T\gtrsim80$ K. 
These results are in good agreement with experimental data~\cite{Scott,Yamada-69,Yamada-2} as shown in Fig.~\ref{fig:Temperature}(a). 
In particular, our simulations reproduce both the temperature evolution of the modes observed experimentally and, more importantly, the temperature ranges, identified by Scott~\cite{Scott-97} , over which the polar and AFD branches cross, which are essential for identifying the temperature regimes of energetic proximity between ARP modes  that favor their coupling. 

To further elucidate the physical consequences of these two temperature regimes, we next examine the evolution with temperature of the roton and of the related dynamical behavior of the ARP modulation obtained from molecular-dynamics simulations. 
As shown in the phonon dispersion curves of Fig.~\ref{fig:Temperature}(b), the polar instabilities at $\Gamma$ are fully suppressed at $T=30$ K. However, due to the energetic proximity between the modes involved in the ARP distortion, their hybridization persists, leading to a flattening of the transverse acoustic branch that is crossed by the in-plane OOR and out-of-plane polar branches and develops a pronounced roton minimum at a finite wavevector $q^\star$. Consequently, the characteristic roton remains clearly visible at temperatures well above the disappearance of the zone-center polar instability, and the associated ARP modulation, characterized by a non-zero $p_z^{q_x}$, is surprisingly preserved in this temperature regime. 
This is confirmed by our lattice dynamics simulations. Increasing progressively the temperature, $P_x^0$ and $P_y^0$ quickly disappear above few Kelvin. However, consistent with the roton character, a coherent state in which the $p_z^{q_x},~\eta_z^{q_x}$ and $\phi_x^{\bar{{q_x}}}$ modes fluctuate synchronously is observed (see Fig.~\ref{fig:Temperature}d,e) leading to a dynamically stabilized coherent state with enhanced lifetime~\cite{Li-25} that persists over the timescale of the simulations, showing non-zero time-averaged amplitudes (see Fig.~\ref{fig:incoherent}).
These results demonstrate that the roton feature and the associated ARP-modulated state do not rely on the presence of $\Gamma$-point polar instabilities. This robustness further suggests that the ARP-modulated state may survive the effects of quantum fluctuations.

As temperature rises, we enter a regime in which in-plane OOR and out-of-plane polar branches gradually separate from the acoustic branch, thereby weakening their coupling (see Fig.~\ref{fig:tdep}). Consequently, the roton minimum hardens, as shown in Fig.~\ref{fig:Temperature}c, while gradually losing its polar character and acquiring a distinct component associated now with rotations about the tetragonal axis that become closer in energy. This evolution is once again reflected in the dynamics: as the roton changes its character when increasing temperature, it no longer promotes the synchronous evolution of the ARP modes producing a loss of correlation at $T\sim60$ K as shown in Fig.~\ref{fig:Temperature}d and Fig.~\ref{fig:incoherent}c. This leads to an incoherent state where fluctuations of the modes that constitute the ARP roton are becoming more erratic, producing the progressive disappearance of the collective state in favor of the $I4/mcm$ phase (see supplemental information Fig.~\ref{fig:tdep}).
\begin{figure*}[ht]
     \centering    \includegraphics[width=\textwidth]{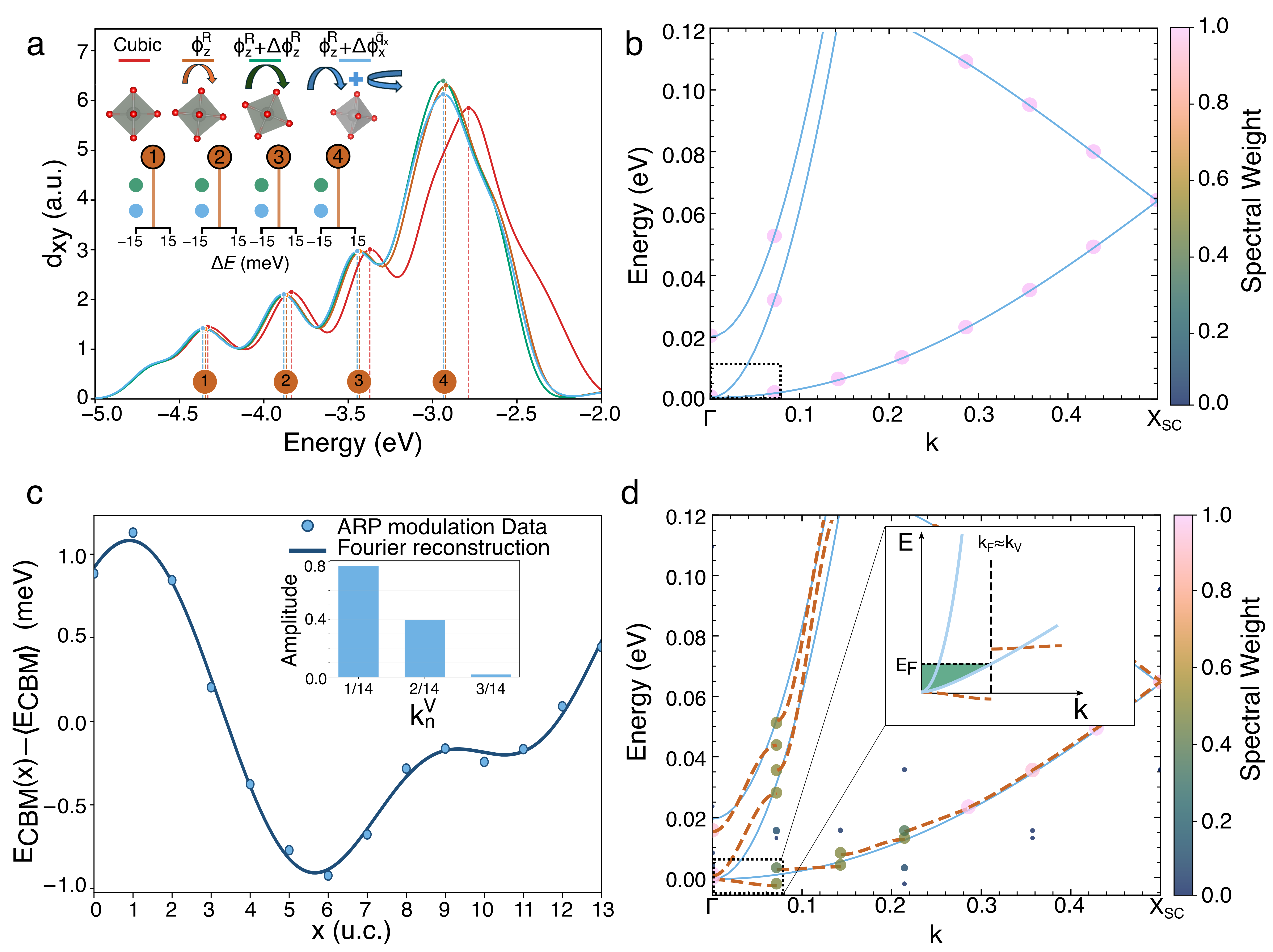} \caption{\textbf{Consequences of the roton modulated state on the low-temperature behaviour of SrTiO$_3$}. \textbf{a,} Projected density of states of the Ti $d_{xy}$ orbital  calculated from first-principles for the cubic reference structure (red), the $I4/mcm$ phase with the equilibrium rotation $\phi_z^R$ (orange), an $I4/mcm$ structure with an enhanced rotation about the same axis,labeled $\phi_z^R+\Delta\phi_z^R$ (green), and the $I4/mcm$ phase with an additional transverse rotation component ($\phi_x^{\bar{{q}_x}}$), labeled  $\phi_z^R+\Delta\phi_{x}^{\bar{{q}_x}}$ (blue). For visualization purposes, the actual amplitudes of $\Delta\phi_z^R$ ($5\%~\phi_z^R$) and $\Delta\phi_{x}^{\bar{{q}_x}}$ ($10\%~\phi_z^R$) have been amplified by a factor of 4 in the reported calculations.   The appearance of transverse rotations (blue) can produce comparable shifts of the crystal-field-related peaks than the enhancement of primary rotations (green) as shown in the inset at a magnified scale for each of the peaks labeled 1-4. \textbf{b,} Lowest conduction bands of the $I4/mcm$ phase using a 40-atom $2\times2\times2$  cell (blue curves), together with the spectral weight of the unfolded bands when using a $14\times2\times2$  $I4/mcm$ supercell, as obtained from first-principles. \textbf{c,} Evolution of the local conduction-band edge along the modulation direction, as computed from first-principles for the $I4/mcm$ phase condensing the ARP distortion, using a $14\times2\times2$ supercell. It points out a clear spatial modulation of the macroscopic potential resulting from the ARP modulation. Inset: Fourier components $k_n^V$ of the modulation of the potential, as obtained from the fit (blue line). \textbf{d,} Reconstructed lowest conduction bands of the $I4/mcm$ phase condensing the ARP distortion (red dashed curves), unfolded in the 40-atom cell of the $I4/mcm$ phase (blue lines). Dots are the spectral weights used for the reconstruction. We see gap openings at k-points compatible with the Fourier component of the modulation. The inset shows a schematic of the band filling, with the $I4/mcm$ bands filled up to the Fermi energy.} 
      \label{fig:electronic} 
\end{figure*}

These results are consistent with experimental observations spanning several decades. 
The anomalous softening of the transverse acoustic branch and the roton feature in bulk SrTiO$_3$ were first anticipated by Müller {\it et al.} ~\cite{Muller-91} and subsequently observed experimentally by Courtens~\cite{Courtens-96} in the 1990s, with the characteristic direction and position of the finite-wavevector anomaly reported at that time amazingly close to our findings. 
As previously discussed, recent finite-$q$ acoustic softening and nanoscale polar textures compatible with our findings have been observed under tensile strain using modern scattering and imaging techniques~\cite{Orenstein-25,Wang-26}.
Together, these observations provide strong experimental evidence for the finite-$q^\star$ ARP instability identified here.

Müller {\it et al.}~\cite{Muller-91} had conjectured the existence of a roton and proposed the formation of a novel quantum-coherent state in order to explain an intriguing anomaly at low temperature in the OOR angle amplitude $\phi_z^R$,  as revealed by electron paramagnetic resonance (EPR). 
In particular, the low-temperature EPR anomaly behaved as  an additional order parameter, leading them to hypothesize that it originates from an unknown finite-$q$ instability.
While our results provide striking evidence together for the emergence of this modulated state and its proposed coherence, the additional OOR transverse components that accompany the ARP modulated state further provide a natural  reinterpretation for the long-standing Müller anomaly.
In addition to the primary OOR  about the tetragonal axis ($\phi_z^R$), the ARP distortion contains indeed smaller modulated OOR ($\phi_x^{\bar{{q}_x}},~\phi_y^{\bar{{q}_x}}$) about the two transverse axes [Fig.~\ref{fig:ModesPDW}c]. As highlighted by our calculated crystal-field splittings (Fig.~\ref{fig:electronic}a), the appearance at low temperature of these additional OOR distortions modify the local crystal-field splitting and produce a shift of the peaks similar to the apparent increase of 5\% in the rotation angle inferred from EPR, without requiring an actual enhancement of the primary $I4/mcm$ rotation $\phi_z^R$. This reinterpretation of experimental data further reconciles M\"uller's anomaly with the view of Scott, who was contesting any enhancement of $\phi_z^R$~\cite{Scott-97,Scott-Iso}.

Beyond its structural signatures, the emergence of the ARP modulated state also has important consequences on the electronic properties of SrTiO$_3$. In the tetragonal $I4/mcm$ phase, the  electronic conduction-band edge is dominated by Ti 3d $t_{2g}$ states, which are split at $\Gamma$ into a doublet and a singlet, as sketched in the electronic dispersion curves of  Fig.~\ref{fig:electronic}b, using a 40-atom cell ($2\times 2\times 2$ pseudocubic supercell). 
The condensation in the $I4/mcm$ phase of the ARP atomic distortion, with a modulation wavelength $\lambda^* = 14$ unit cells, produces the appearance of extra Fourier components in the external potential seen by the electrons, at characteristic  wavevectors  $k^V_n= (\frac{n}{14},0,0)$ of the simple cubic Brillouin zone. This can be traced back and quantified from the evolution of the local conduction band edge of the modulated structure, along the supercell, as illustrated in Fig.~\ref{fig:electronic}c.
These additional Fourier components of the potential are then expected to open gaps in the electronic dispersion curves at $k= (\frac{n}{28},0,0)$ of the simple cubic Brillouin zone or at $k_{SC}= (\frac{n}{14},0,0)$, considering instead a 40-atom supercell. This is nicely confirmed in Fig.~\ref{fig:electronic}d reporting the computed electronic band structure of the $I4/mcm$ phase condensing the ARP distortion ($14\times 2 \times 2$ modulated structure), unfolded on a 40 atom supercell, for direct comparison with the reference $I4/mcm$ phase:
it clearly reveals the appearance of a series of consecutive gaps, the amplitudes of which are decreasing consistently with those of the related Fourier components $k^V_n$ of the modulation potential.
The electronic band reconstruction testifies of a strong electron-roton coupling and provides a visual and direct electronic signature of the modulated state.

In the undoped insulating state, gap openings around the conduction band edge have no direct energetic consequence, despite the strong electron-phonon coupling associated with the modulation. Upon electron doping, however, the situation changes substantially. As illustrated in the inset of Fig.~\ref{fig:electronic}d, when doping brings the Fermi wavevector $k_F$ close to the wavevector $k_1^V$ at which the band gap opens, the effects of the electron-phonon coupling are strongly exacerbated, as the ARP modulation now directly affects occupied conduction states, lowering their energy and thereby enhancing the energetic gain associated with the distortion. Thus, although electron doping is known to suppress the ferroelectric instability~\cite{Tsymbal-12,Annette-20}, it is expected instead here to stabilize the ARP distortion through a mechanism reminiscent of the Peierls instability~\cite{Peierls1955}.
Strikingly, this condition corresponds to a carrier density of $n \sim 10^{19}$ cm$^{-3}$, for which $k_{\rm F}\approx 1/33$ u.c.$^{-1}$, assuming an anisotropic parabolic dispersion with effective masses obtained from first-principles calculations. This points out that the ARP roton $q^\star$ can couple effectively with electronic states at carrier densities where superconductivity is experimentally observed~\cite{Triscone-20,Schooley-64}.

This scenario is further supported by our estimate of the electron-roton coupling strength and the resulting superconducting transition temperature within the Migdal–Eliashberg framework using the double-delta approximation~\cite{Migdal,Eliashberg-60,Giustino}. 
As detailed in the Supplementary information, our calculations yield a roton coupling constant of $\lambda=0.4$, corresponding to superconducting transition temperatures of $T_c\simeq0.54$ K -- $0.12$ K for screened Coulomb parameters $\mu^\star=0.10$ -- $0.15$, respectively, within the McMillan–Allen–Dynes formalism~\cite{McMillan,Allen-Dynes}. These values are in amazing agreement with the superconducting transition temperatures experimentally observed in SrTiO$_3$ at similar carrier densities~\cite{Koonce-67} further supporting a possible role of the $q^\star$-ARP roton in the microscopic origin of superconductivity in the dilute-carrier regime.


In conclusion, we have shown that the anomalous low-temperature behavior of SrTiO$_3$ can be understood through the emergence of an intrinsic and composite ARP atomic modulation, stabilized by anharmonic coupling between phonons of the acoustic, antiferrodistortive (OOR) and polar branches. The strong hybridization of these modes  in the low-temperature regime produces a roton finite-$q^\star$ minimum in the phonon dispersion curves, that is responsible of the modulation.
At $T=0$ K, the cooperative anharmonic interactions between the modes lower the energy enough to stabilize the ARP modulated state below the  avoided polar ground states of SrTiO$_3$. 
At low temperatures, the same couplings, granting a mixed ARP character to the roton feature, induce a coherent state where the modes oscillate synchronously, acquiring an enhanced lifetime and showing dynamical stability. On heating, the ARP modes separate, yielding progressive disappearance of the modulated state in favor of the $I4/mcm$ phase.

This roton-based framework provides a unified microscopic explanation to independent long-standing puzzles of the low-temperature physics of SrTiO$_3$, including the emergence of modulated polar order, the Müller anomaly of OOR, or the emergence of superconductivity at ultra-low carrier densities. 
What makes SrTiO$_3$ so unique is likely that, consistently with its Goldschmitd tolerance factor~\cite{Goldschmidt1926} close to 1, it shows weak polar and OOR competing instabilities, yielding ultimately the crossing between relevant acoustic, OOR and polar phonon branches happening in the low-temperature regime. But our findings are not restricted to SrTiO$_3$ and points toward a more global mechanism. In displacive perovskites, the softening of distinct phonon branches will unavoidably affect band crossings with temperature, potentially giving rise to analogous instabilities and favor spatially modulated configurations in substantially different temperature regimes. This raises the possibility that precursor states associated with incommensurate order are more widespread than currently appreciated, and highlights the importance of considering coupled low-energy degrees of freedom for understanding modulated phases in complex materials. Tracking roton-like features in the phonon dispersion curves in those compounds might help identifying and rationalizing the emergence of such modulated phases.

\backmatter

\bmhead{Acknowledgements}
We acknowledge Annette Bussmann-Holder, Javier Junquera, Jean-Marc Triscone and Julien Varignon for usuful discussions. F.G.O. acknowledges financial support from MSCA-PF 101148906 funded by the European Union and the Fonds de la Recherche Scientifique (FNRS) through the grant FNRS-CR 1.B.227.25F and the Consortium des Équipements de Calcul Intensif (CÉCI), funded by the F.R.S.-FNRS under Grant No. 2.5020.11 and the Tier-1 Lucia supercomputer of the Walloon Region, infrastructure funded by the Walloon Region under the grant agreement No. 1910247. Ph. G. and L.B. also acknowledge support from the Fonds de la Recherche Scientifique (FNRS) through the PDR project TOPOTEX (Grant No. T.0128.25).

\bmhead{Author contributions}
F.G.O and Ph.G. conceptualized the work. Ph. G. supervised the research and led the project. F.G.O. performed the FP relaxations, electronic structure calculations and SP molecular dynamics simulations, and analyzed the results. L.B. developed the SP model, performed the DFPT phonon calculations, and carried out the SP molecular dynamics simulations and TDEP analysis of the finite-temperature phonons. F.G.O and Ph.G. rationalized the results, unified the concept and wrote the manuscript. All authors contributed to the revision and editing of the manuscript and approved the final version.

\bmhead{Competing interests}
The authors declare no competing interests.

\section*{Methods}
\emph{Second-principles model construction.-}
Second-principles (SP) atomistic models were constructed using the \textsc{Multibinit}~\cite{abinit26} software by fitting data from density functional theory (DFT) calculations produced with the \textsc{Abinit}~\cite{abinit26} software package. The generalized gradient approximation (GGA) with the PBESol exchange-correlation functional and a plane wave pseudopotential approach with optimized norm-conserving pseudopotentials from the PseudoDojo server~\cite{hamann2013optimized,van2018pseudodojo} were employed, considering as valence electrons $4s^{2}4p^65s^2$ for Sr, $3s^23p^63d^24s^2$ for Ti, and $2s^22p^4$ for O. A plane wave energy cutoff of $65\ Ha$ and an $8\times 8\times 8$ $\Gamma$-centered k-point mesh were used. 

The harmonic part of the models was derived from Density Functional Perturbation Theory (DFPT) as implemented in \textsc{Abinit}~\cite{abinit26}. Dynamical matrices for the relaxed cubic $Pm\bar{3}m$ structure were computed using a $8\times 8\times 8$ $q$-point mesh. Various properties such as dipole-dipole interaction, Born effective charge, strain-phonon coupling, and elastic constants were extracted from the DFPT framework.
The anharmonic part of the SP potential was fitted from DFT configurations. A cutoff radius of $\sqrt{3}/2$ times the cubic lattice cell parameter was used to generate the anharmonic symmetry-adapted terms (SAT) from third to eighth order, considering strain-phonon coupling and anharmonic elastic constants. The model has been extensively tested and used in previous works~\cite{GomezOrtiz26}, and is publicly available along with a validation of its accuracy following Ref.~\cite{model}.

\emph{Tempearture renormalized phonons.-} 
Temperature-dependent phonon properties were calculated using the Temperature-Dependent Effective Potential~\cite{Hellman-13,Knoop-24} method, which has demonstrated good accuracy in reproducing experimentally measured phonon spectra in SrTiO$_3$~\cite{Xing-20}.
The lattice anharmonic dynamics were performed on a $4\times 4 \times 4$ supercell of the five-atom cubic perovskite cell (320 atoms). The equilibrium lattice parameters at each temperature were first obtained from hybrid Monte Carlo simulations in the NPT ensemble at zero pressure. The cell was then held fixed at these parameters and a second hybrid Monte Carlo simulation was run in the  NVT ensemble, from which temperature-dependent effective higher-order interatomic force constants were extracted using a pair cutoff radius of 7~\AA{}, which is necessary to capture the contributions of more distant neighbors required for the development of the roton minimum.
The phonon eigendisplacements were subsequently analyzed using {\sc{agate}} software~\cite{agate}. 

\emph{Structural relaxations and finite temperature simulations.-} Both, first- and second-principles structural relaxations were carried out using the Broyden-Fletcher-Goldfarb-Shanno (BFGS) algorithm as implemented in {\sc{abinit}}~\cite{abinit26}. SP finite-temperature molecular dynamics simulations were performed by first thermalizing the system in the canonical (NVT) ensemble for 5 ps, followed by 10 ps of simulations in the isothermal–isobaric (NPT) ensemble. The NPT trajectories were used to compute thermal averages and to analyze the temperature dependence and dynamical evolution of the distortions.

To further validate the SP predictions, additional first-principles structural relaxations were performed for the modulated state. These calculations employed a $\lambda\times2\times2$ supercell, with $\lambda$ systematically varied from 8 to 18 u.c. A plane wave energy cutoff of 40 Ha and a $1\times4\times4$ $k$-point mesh were used. The structures were fully relaxed until the forces and stresses were converged to $5\times10^{-5}$ Ha/bohr and $1\times10^{-7}$ Ha/bohr$^3$, respectively. Starting from the atomic coordinates obtained from the SP calculations, the relaxations converged to metastable ARP modulation configurations for all considered $\lambda$, retaining similar polar, acoustic, and rotational mode amplitudes.

\emph{Electronic structure calculations.-} The projected density of states was calculated using the {\sc siesta}~\cite{siesta} code. The one-electron Kohn–Sham eigenstates were expanded in a
basis of strictly localized numerical atomic orbitals. For O, the 2s and 2p states are described with double-$\zeta$ precision complemented by a single-$\zeta$ 3d polarization shell. For Ti, double-$\zeta$ functionsare used for the 4s and 3d states. For Sr, the 5s state is described by double-$\zeta$ functions while single-$\zeta$ 5p and 4d shells are included to provide additional angular flexibility. All the parameters that define the range and the
shape of the atomic orbitals have been optimized variationally in
bulk cubic centrosymmetric SrTiO$_3$ following the recipe given in~\cite{siestabasis}.
The electronic density, Hartree, and exchange-correlation potentials, as well as the corresponding matrix elements between
the basis orbitals, were calculated in a uniform real-space grid. An equivalent plane wave cutoff of 1200 Ry was used to represent the charge density.

Finally, electronic bands were computed with {\sc{abinit}} by condensing the modulated distortions characterizing the ARPM state in a 14$\times$2$\times$2 supercell and folded back to the conventional 2$\times$2$\times$2 cell of the tetragonal $I4/mcm$ phase using the \texttt{fold2bloch} utility distributed within {\sc{abinit}}\cite{abinit26}.

\clearpage\onecolumn
\setcounter{figure}{0}
\renewcommand{\figurename}{Fig.}
\renewcommand{\thefigure}{S\arabic{figure}}
\bmhead{Supplementary information}
\bmhead{DFT relaxations}
Both, DFT and second-principles calculations result on a energetic preference for a finite-wavelength ARP modulation. However, the position of the predicted optimal periodicity differs between the two descriptions as shown in Fig.~\ref{fig:instability_phonons}. The second-principles model exhibits its minimum at $\lambda=18$ u.c., whereas the DFT calculations favor a slightly shorter modulation period of approximately $\lambda=14$ u.c. 
This difference reflects the quantitative sensitivity of the long-wavelength energy landscape to the underlying first-principles energetics, while the existence of a finite-$q$ minimum and the character of the associated distortion are robust. In particular, the second principles-relaxed structures exhibit the same characteristic modulations identified in the main text and shown in Fig.~\ref{fig:ModesPDW} with comparable mode amplitudes.
Importantly, both the DFT as well as the second principles-relaxed ARP modulations are lower in energy than the competing $I4/mcm$ and $Cc$ structures, establishing the modulated state as the thermodynamical ground state within both levels of theory.

DFT energies in Fig.~\ref{fig:instability_phonons} are reported with their estimated uncertainty. The error bars account for the use of different supercell sizes, $\lambda\times2\times2$ with $\lambda=8,10,12,14,16,$ and $18$, while keeping the same $1\times4\times4$ $k$-point mesh. We define the uncertainty as the maximum deviation between the energy of the $Cc$ reference phase calculated for each supercell periodicity and that obtained using the 40-atoms reference supercell.

\bmhead{Symmetry considerations for the acousto-roto-polar coupling} In the main text, we discussed the coupling between a transverse acoustic mode, a transverse optical (polar) mode, and an octahedral rotational mode. A necessary condition for such a coupling to be symmetry allowed is that the direct product of the irreducible representations (irreps) of the three modes contains the totally symmetric representation, $\Gamma_1$. In the present case, the relevant modes are the polar mode polarized along the principal axis of the tetragonal distortion of the $I4/mcm$ phase, i.e., along $z$, denoted by $p_z^{q_x}$,the octahedral rotational modes about the two perpendicular directions $x$ and $y$, denoted $\phi_x^{\bar{{q_x}}}$, $\phi_y^{\bar{{q_x}}}$ and the transversal acoustic mode $\eta_z^{q_x}$.
As discussed by Scott and Ledbetter~\cite{Scott-97}, in tetragonal SrTiO$_3$ the transverse acoustic mode transforms according to the  $E_u$ irrep, while the polar mode polarized along the tetragonal axis transforms according to $A_{2u}$. The octahedral rotations about the two in-plane axes form an $E_g$ doublet. The corresponding product is therefore symmetry allowed since $E_g \otimes A_{2u} \otimes E_u \supset \Gamma_1$.

Therefore, the coupling between the polar, rotational, and acoustic modes is not accidental but follows directly from the symmetry of the tetragonal phase. At low temperatures, where these modes have comparable characteristic frequencies, this symmetry-allowed interaction can promote their coupling and contribute to the formation of the acousto-roto-polar modulated state discussed in the main text.

\bmhead{Second-principles molecular dynamics}
To study the temperature evolution of the acousto-roto-polar state we performed molecular dynamics simulations at different temperatures as explained in the methods section. 
\begin{figure}[h]
     \centering
      \includegraphics[width=\columnwidth]{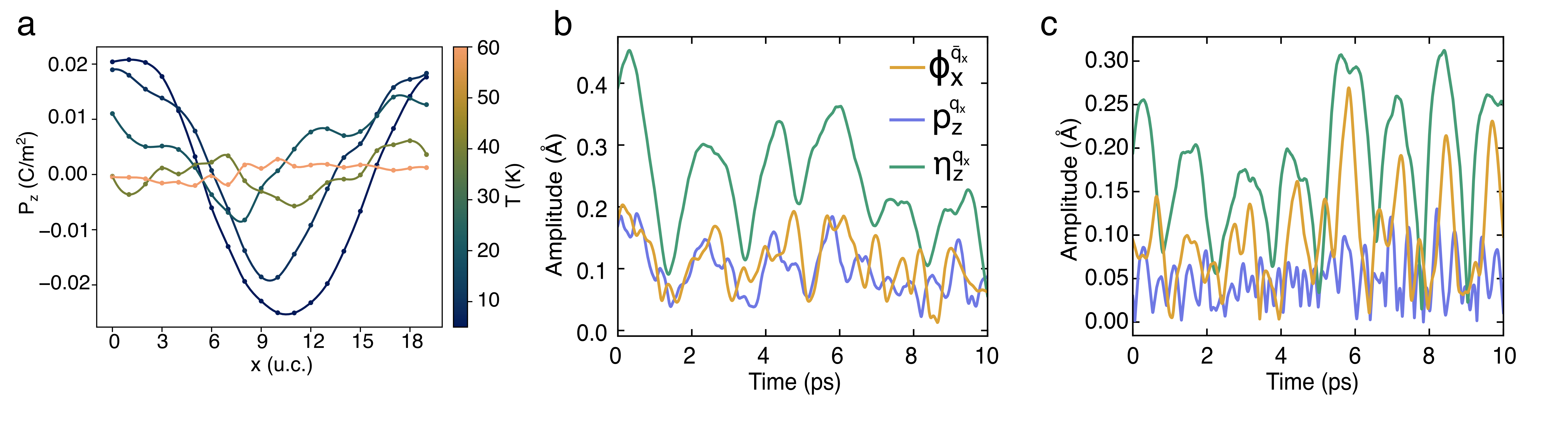}
      \caption{\textbf{Temperature evolution of the acousto-roto-polar modulated state}. \textbf{a,} Temperature dependence of the modulated polarization as predicted by second-principles simulations. The polarization profile at each temperature corresponds to the average state over the 10 ps molecular-dynamics trajectory. \textbf{b, c,} Evolution of the different distortion modes during the dynamics at $T=20$ and $T=65$ K respectively, as predicted by second-principles simulations. } 
      \label{fig:incoherent} 
\end{figure}
As discussed on the main text Fig.~\ref{fig:Temperature}, increasing temperature progressively suppresses the coupling between the out-of-plane polar, in-plane rotational, and acoustic modes composing the modulated state which translates in a loss of coherence between the temporal fluctuation of these modes. This loss of coherence is directly visible in the time evolution of the corresponding mode amplitudes, shown in Fig.\ref{fig:incoherent}b,c. At low temperature $T=20$ K, the three modes evolve synchronously over extended time intervals, preserving the characteristic modulation of the ARP state Fig.\ref{fig:incoherent}a. With increasing temperature, their relative phase coherence is progressively lost, with each mode fluctuating increasingly independently of the others, such that the collective state becomes more transient and eventually disappears from the time-averaged structure.

Accordingly, the time-averaged structure evolves from the modulated acousto-roto-polar state at low temperatures ($\sim T\leq 40$ K) toward the tetragonal $I4/mcm$ structure at higher temperatures, where the finite-$q$ modulation of the polarization is no longer retained as shown in Fig.\ref{fig:incoherent}a at $T=60$ K.

\bmhead{Temperature dependence of the phonon spectrum}
To investigate the evolution of the roton minimum in the acoustic branch of SrTiO$_3$ with temperature, we calculated the temperature-dependent phonon spectrum using the temperature-dependent effective potential ({\sc{tdep}}) method~\cite{Hellman-13,Knoop-24} as explained in the methods section. 
\begin{figure}[h]
     \centering
      \includegraphics[width=\columnwidth]{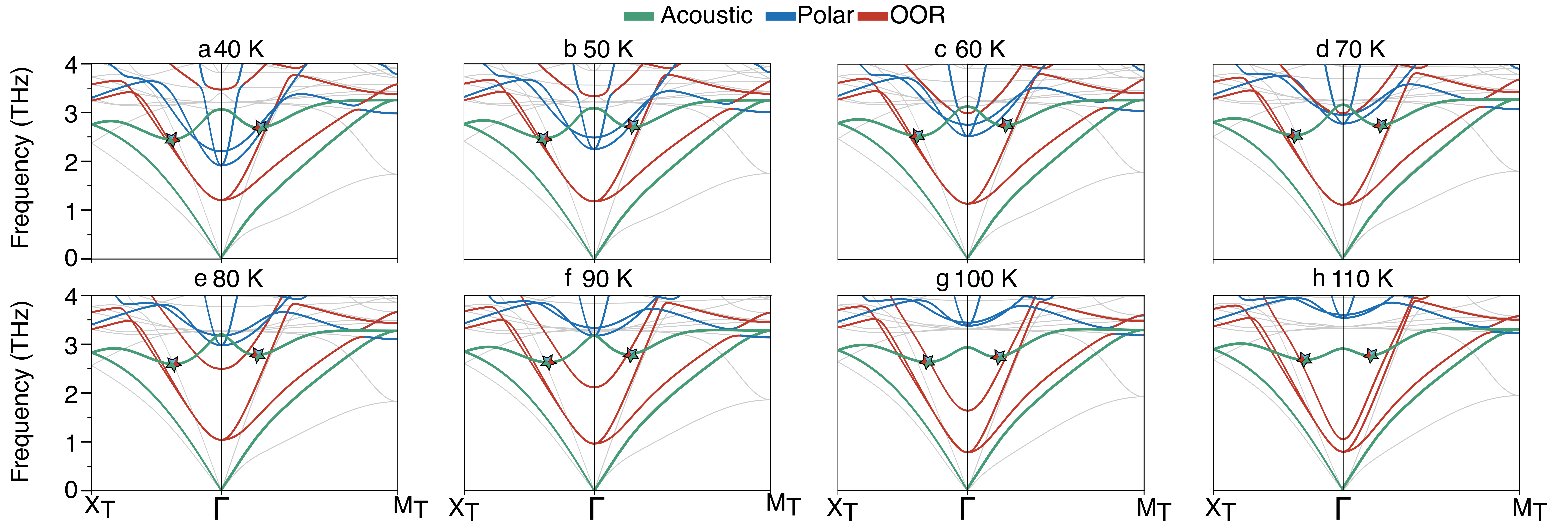}
      \caption{\textbf{Temperature evolution of the phonon dispersion curves}. Temperature-dependent phonon dispersions of SrTiO$_3$ in the 20 atom $I4/mcm$ phase along the $X_T-\Gamma-M_T$ path at \textbf{a,} 40K, \textbf{b,} 50K, \textbf{c,} 60K, \textbf{d,} 70K, \textbf{e,} 80K, \textbf{f,} 90K, \textbf{g,} 100K and \textbf{h,} 110K. The acoustic branch presenting the roton, the polar branches and the antiferrodistortive branches are highlighted as indicated in the legend.} 
      \label{fig:tdep} 
\end{figure}

As shown in Fig.~\ref{fig:tdep}, upon cooling, the polar mode softens toward the acoustic branches, while the AFD branches hardens resulting in strong hybridization with the acoustic branch. As a result, the transverse acoustic branch adopt a roton minimum that evolves with temperature. The evolution with temperature of these frequencies is shown in Fig.~\ref{fig:Temperature}c alongside experimental evidence. 

As discussed in the main text, depending on the character of the polar and AFD branches that cross the acoustic one at each temperature the character of the roton evolves. At lower temperatures where out-of-plane polar and in-plane rotational branches are close to the acoustic branch the roton has the character discussed in Fig~\ref{fig:ModesPDW}. At higher temperatures, however, the polar character progressively disappears and the rotational character acquires a out-of-plane component rather than an in-plane component. 

Moreover, as discussed in the main text, the roton feature in the acoustic branch is well reproduced throughout the 
$[001]_{pc}$ plane, and is therefore not restricted to the $[100]_{pc}$ direction but can also arise along diagonal $[110]_{pc}$ ones as shown in Fig.~\ref{fig:tdep} along the $X_T-\Gamma-M_T$ path.

\begin{figure}[h]
     \centering
      \includegraphics[width=10cm]{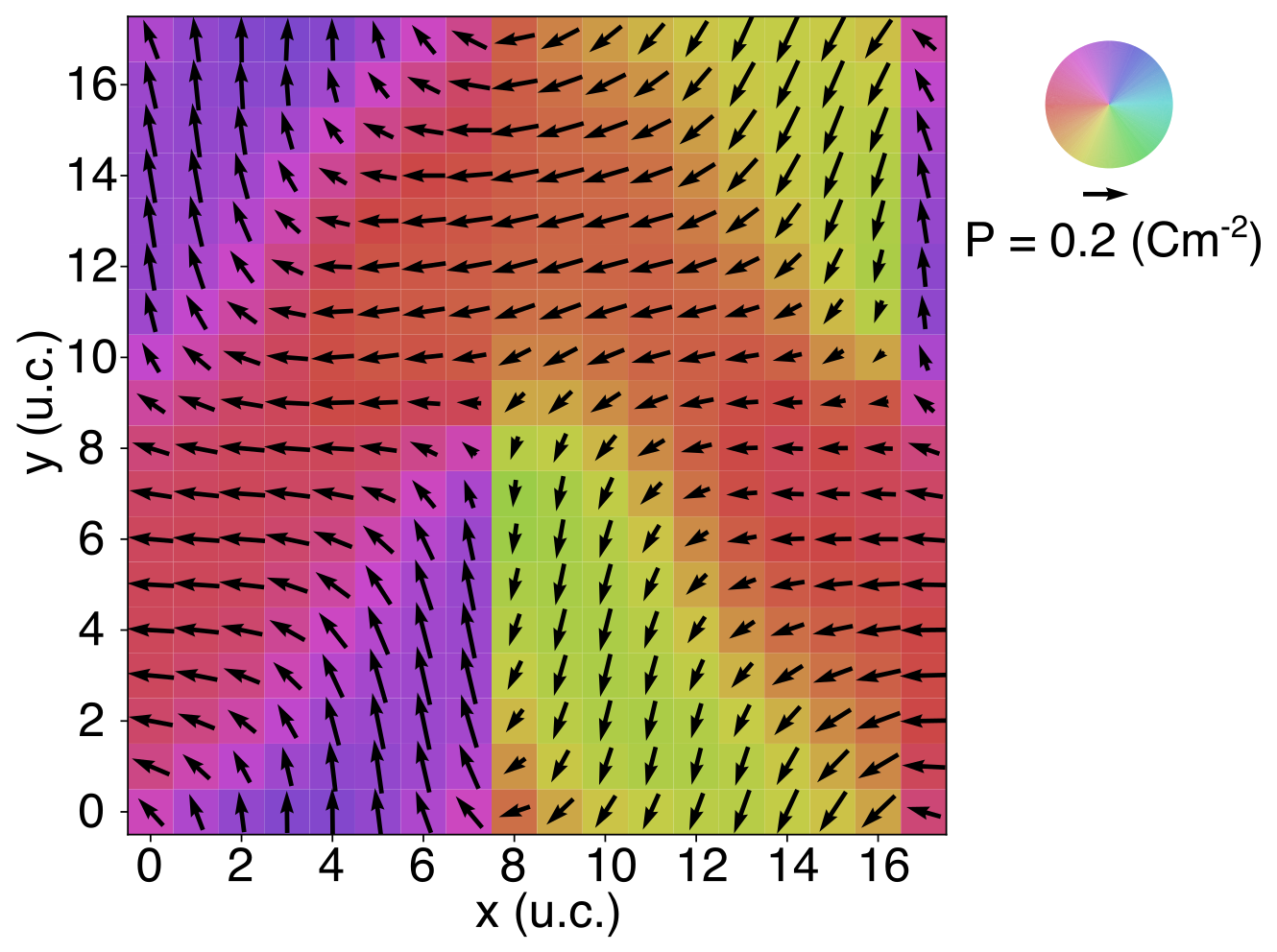}
      \caption{\textbf{Relaxed polar nano-regions}. Polarization map obtained from our second-principles model, showing a fully relaxed modulated polar state in SrTiO$_3$. Each arrow represents the polarization vector within a unit cell, with its direction indicated by the color scale and its magnitude represented by the arrow length.} 
      \label{fig:nanoregions} 
\end{figure}

Interestingly, the condensation of these modes along different directions yields comparable energy lowerings and promotes the modulation of the polarization. It can generate polar nano-regions resembling those recently observed experimentally~\cite{Zhang-26}, as shown in Fig.~\ref{fig:nanoregions} after full structural relaxation. Both the characteristic dipole pattern and the spatial extent of these modulations are consistent with the experimental observations, which reveal stable polar nano-regions at $T=69$ K~\cite{Zhang-26} extending over approximately 25 unit cells.
\bmhead{Electron-phonon coupling}
To assess the potential role of the ARP modulation in superconductivity, we estimate the electron–phonon coupling strength and the corresponding superconducting transition temperature within the McMillan–Allen–Dynes formalism as described in the main text. 
Given the computational cost of treating the large supercell associated with the ARP modulation, we estimate the electron–phonon coupling strength using a finite-difference frozen-phonon approach~\cite{Julien-1,Julien-2}. Within the double-delta approximation~\cite{Migdal,Eliashberg-60,Giustino}, the electron–phonon coupling constant is estimated as
\begin{equation}
    \lambda = N(\varepsilon_F)\dfrac{\Delta E^2 }{\omega^2\Delta Q^2}\nonumber,
\end{equation}
where $N(\varepsilon_F)$is the electronic density of states at the Fermi level, $\Delta E$ is the gap induced by the distortion, $\omega$ is the characteristic frequency of the roton mode, and $\Delta Q$ is its mass-normalized distortion amplitude.
We estimate $N(\varepsilon_F)$ using a parabolic-band approximation at a carrier density of $n \sim 10^{19}$ cm$^{-3}$ with effective masses computed within first-principles. Then, using $\omega \sim 90$ cm$^{-1}$ as the characteristic frequency of the roton mode, $\Delta E = 6$ meV as the induced gap associated with the condensation of the distortion with a mass-normalized amplitude of $\Delta Q = 0.573$~\AA$\sqrt{\mathrm{amu}}$, we obtain an electron–phonon coupling constant of $\lambda=0.4$. We then estimate the superconducting transition temperature using the McMillan–Allen–Dynes formalism~\cite{McMillan,Allen-Dynes}. For screened Coulomb pseudopotentials of $\mu^\star=0.1$ and $0.15$, we obtain $T_c\simeq0.54$ K and $0.12$ K, respectively as reported in the main text.
\end{document}